\documentclass[preprint,12pt]{elsarticle}
\usepackage{amsfonts,amsmath, amssymb, dsfont, eucal, natbib}
\usepackage{graphicx, color}
\usepackage{hyperref}
\usepackage{epsfig}
\def\sH{\mathcal H}\def\sK{\mathcal K}\def\sE{\mathcal E}\def\sS{\mathcal S}
\def\Tr{\mathrm {Tr}}\def\<{\langle}\def\>{\rangle}

\def\Lin{\mathcal L}
\def\Rng{\mathrm{Rng}}
\def\nor#1{|\!| #1 |\!|}
\newtheorem{theorem}{Theorem}

\begin{document}

\begin{frontmatter}

\title{Adaptive Bayesian and frequentist data processing for quantum tomography}

\author[quit,infn]{Giacomo M. D'Ariano} 
\author[quit,cnism]{Daniele F. Magnani}
\author[quit,infn]{Paolo Perinotti}

\address[quit]{QUIT Group, Dipartimento di Fisica "A. Volta'', via
  Bassi 6, 27100 Pavia, Italy}
\address[infn]{Istituto Nazionale di Fisica Nucleare (INFN), Sezione
  di Pavia, via Bassi 6, 27100 Pavia, Italy}
\address[cnism]{Consorzio Nazionale Interuniversitario per le Scienze
  Fisiche della Materia (CNISM), unit\`a di Pavia, via Bassi 6, 27100
  Pavia, Italy} 

\begin{abstract}
  The outcome statistics of an {\em informationally complete} quantum
  measurement for a system in a given state can be used to evaluate
  the ensemble expectation of any linear operator in the same state,
  by averaging a function of the outcomes that depends on the specific
  operator. Here we introduce two novel data-processing strategies,
  non-linear in the frequencies, which lead to faster convergence to
  theoretical expectations.
\end{abstract}

\begin{keyword}
quantum estimation \sep quantum tomography\textit{}
\PACS 03.65.Wj \sep 03.65.-w \sep 03.67.-a
\end{keyword}
\end{frontmatter}

\date{\today}

\section{Introduction}
In Quantum Mechanics measuring a single observable provides only
partial information about the state of the measured system. According
to the Born interpretation, the quantum state is a rule for evaluating
the outcome probabilities in all conceivable measurements, and a
complete information about the quantum state requires a thorough
outcome statistics for a {\em quorum} of observables, or for a
suitable {\em informationally complete measurement} (shortly
info-complete)\cite{prugo,bush}, in conjunction with a suitable
data-processing, as it is done in quantum tomography (for a review see
Ref. \cite{tomo}).  There are two main classes of approaches in
quantum tomography: a) averaging ''patterns functions'' a method
initiated in Ref. \cite{pattern}; b) Maximum likelihood techniques
\cite{like}\par
Method a) has the advantage of providing any expectation value, e.g. a
single density matrix element, without the need of estimating the
entire density operator. However, the estimated full matrix is not
necessarily positive, which is not a serious drawback, since the non
positivity falls within a small fluctuation for large numbers of
data.\par
Method b) has the advantage of providing a positive density operator,
with smaller fluctuations, however, it has the more serious drawback
of needing to estimate the full density matrix, while is exponentially
large versus the number of systems, and, in the infinite dimensional
case needs a dimensionality cutoff which introduce a bias that is
under control only if there is some prior knowledge of the state.\par
In a recent paper \cite{optproc} the optimal data-processing for
evaluating ensemble averages from experimental outcomes was derived
for a completely general setting within a Bayesian scheme that assumes
a prior probability distribution of states. Using as optimality
criterion the rate of estimated-to-theoretical convergence of
averages, the optimal data-processing itself depends on the prior
distribution of states.

The purpose of the present paper is to exploit the dependence of the optimal data-processing on the
prior distribution of states, in order to improve the convergence rate using an adaptive
data-processing scheme. We will consider info-complete measurements---more generally than a quorum
of observables---whose statistics allows to reconstruct all possible ensemble averages. Estimation
of the quantum state itself is equivalent to the estimation of all possible ensemble averages. We
will adopt the natural figure of merit used in Ref. \cite{optproc}, which, in the present context,
represents the estimated-to-theoretical convergence rate (in Hilbert-Schmidt distance) of the state.
As we will see, exploiting the dependence of the optimal data-processing on the prior state leads to
two different data processing strategies, which both improve the convergence rate compared to the
standard tomographic procedures, and are easily implementable and computationally efficient:
\begin{itemize}
\item[] Method 1 (Bayesian iterative procedure): Bayesian update of the prior distribution after the
  first state reconstruction, with subsequent iteration of the optimization. 
\item[] Method 2 (Frequentist approach):
replace the theoretical probability distribution of the info-complete in the optimal data-processing
with the experimental frequencies. 
\end{itemize}
We will see that numerical simulations carried out with both methods show relevant improvement of
convergence compared to the plain non adaptive processing of Ref.  \cite{optproc}.

The paper is organized as follows. In Section \ref{opti} we re-derive the optimal data-processing
for given prior distribution of Ref. \cite{optproc} within an improved theoretical framework. In
Sections \ref{bay} and \ref{freq} we introduce Methods 1 and 2, respectively.  Finally, in Section
\ref{exa}, we present numerical simulations for testing both methods in comparison with the original
plain non adaptive data-processing, and in Section \ref{conc} we end the paper with concluding remarks.

\section{Optimization of the data processing}\label{opti}

In the modern formulation of Quantum Mechanics, the state of a quantum
system associated to a $d$-dimensional Hilbert space $\sH\sim{\mathbb
  C}^d$ is represented by a {\em density matrix}, namely a positive
operator $\rho\geq0$ with $\Tr[\rho]=1$. The Born formula provides the
probabilities of outcomes in a quantum measurement in terms of the
state $\rho$ as follows
\begin{equation}
  p(i|\rho):=\Tr[\rho P_i],
\label{eq:Born}
\end{equation}
where the POVM $(P_i)$ (Positive Operator Valued Measure) is a set of
(generally non orthogonal) positive operators $P_i \geq 0$ resolving the identity as $\sum_{i=1}^N
P_i= I$, thus guaranteeing positivity and normalization of probabilities.

The linear span of the POVM elements $P_i$, defined as $\sS:=
\text{Span} \{P_i \}_{1 \leqslant i \leqslant n}$, is a linear
subspace of the space $\Lin(\sH)$ of linear operators on $\sH$,
and we will take as a canonical basis in $\Lin(\sH)$ the
operators $|m\>\<n|$, where $|n\>$ is an orthonormal basis thus representing operators $X$ by the vectors
of their matrix elements $X_{m,n}=\<m|x|n\>$. A POVM is
{\em info-complete} if $\sS\equiv\Lin(\sH)$, namely
all operators $X\in\Lin(\sH)$ can be expanded on the POVM
elements, and it is possible to determine all ensemble averages
$\<X\>_\rho$, as in Quantum Tomography. For each complex operator $X
\in\sS $ the following decomposition holds
\begin{equation} \label{eq:decomposition} X = \sum_{i=1}^N
  f_i[X]\,P_i,
\end{equation}
where $f_i[X]$ is not unique if the set $\{P_i \}$ is over-complete.

With the above expressions we can write the ensemble average of $X$ as follows:
\begin{equation}\label{eq:average} 
  \<X\>_\rho := \Tr[X \rho]=\sum_{i=1}^N f_i[X] p(i |\rho),
\end{equation}
with the following statistical error
\begin{equation}
  \left( \delta X ^{2} \right)_{\rho} := \sum_{i=1}^N | f_i[X] |^{2} p(i |\rho) - | \< X \>_\rho |^{2}
\end{equation}

In a Bayesian scheme one has an \emph{a priori} ensemble $\sE
:= \{ \rho_{i},p_{i} \}$ of possible states $\rho_{i}$ of the
quantum system occurring with probability $p_{i}$. We want to minimize
the average statistical error on $\mathcal E$ in the determination of
the expectation value of $X$, namely the variance
\begin{equation}\label{eq:variance}
  \left( \delta X ^{2} \right)_{\varepsilon}:= \sum_{i=1}^N | f_i[X] |^{2} p(i |\rho_{\varepsilon}) - \overline{| \< X \> |^{2}
  }_{\varepsilon},
\end{equation}
where $\rho_{\varepsilon} = \sum_{i} p_{i}\rho_{i}$ and $\overline{|
  \< X \> |^{2}}_{\varepsilon} =\sum_{i} p_{i}|\Tr[\rho_{i}X]|^{2}$ is
the squared modulus of the expectation of $X$ averaged over the states
in the ensemble (since this term depends only on the ensemble it will
be neglected from now on). Using Eq.(\ref{eq:Born}) the first term
in Eq.(\ref{eq:variance}) can be rewritten as
\begin{equation}\label{eq:relevantpart}
  \Sigma_{f} (X) := \sum_{i=1}^N |f_i[X]|^{2} \Tr[P_{i} \rho_{\varepsilon}].
\end{equation}

Given a POVM $(P_i)$, it is possible to define a linear map $\Lambda$
from an abstract N-dimensional space $\sK$ of coefficient
vectors ${\bf c}\in\sK$ to $\Lin(\sH)$, with range $\sS$:
\begin{equation}
  \Lambda {\bf c} = \sum_{i=1}^{N} c_{i} P_{i},
\end{equation}
so that using the canonical basis in $\sK$, $\Lambda$ has matrix elements
$\Lambda_{mn,i}=(P_{i})_{mn}$. A generalized inverse (shortly g-inverse) of $\Lambda$ is any matrix
$\Gamma$ representing linear operators from $\Lin(\sH)$ to $\sK$ such that the following identity
holds
\begin{equation}\label{g-inv}
  \Lambda \Gamma \Lambda=\Lambda
\end{equation}
Notice that the matrix elements $(\Gamma_{i,mn})$ of $\Gamma$ define a
set of operators $D_i$ with matrix elements $(D_{i})_{mn}:=
\Gamma_{i,mn}^*$. The role of g-inverse $\Gamma$ is assessed by the two following important 
theorems
\begin{theorem}\label{tbap1}
The following statements are equivalent
\begin{enumerate}
\item{$\Gamma$ is a g-inverse of $\Lambda$}
\item{For all ${\bf y}\in\mathrm{Rng} (\Lambda)$, ${\bf x}=\Gamma{\bf y}$ is a solution of
    the equation $\Lambda{\bf x}={\bf y}$.}
\end{enumerate}
\end{theorem}
{\bf Proof.} See Ref. \cite{bapat}.
\begin{theorem}\label{tbap2}
For all g-inverse $\Gamma$ of $\Lambda$ all solutions of $\Lambda{\bf x}={\bf y}$ are of the form
\begin{equation}
  {\bf x} = \Gamma{\bf y}+(I-\Gamma\Lambda){\bf z}, 
  \label{gensol}
\end{equation}
with arbitrary ${\bf z}$.
\end{theorem}
{\bf Proof.} See Ref. \cite{bapat}.

\medskip
We now define a norm in $\sK$ as follows
\begin{equation}\label{eq:geninverse}
  \nor{\bf c}^2_\pi:=\sum_{i=1}^N |c_i|^2\pi_{ii},
\end{equation}
where $\pi_{ij}=\delta_{ij}\pi_{ii}$ is a positive matrix which is
diagonal in the canonical basis in $\sK$. In terms of $\pi$ we define the {\em minimum norm}
g-inverses $\Gamma$ that satisfy \cite{pseudoinverse}
\begin{equation}\
  \pi \Gamma\Lambda=\Lambda^\dag \Gamma^\dag\pi.
\label{minnorm}
\end{equation}
Notice that the present definition of minimum norm g-inverse requires that the norm is induced by a
scalar product (in our case $\vec a\cdot\vec b:=\sum_{i=1}^N a_i^*\pi_{ii}b_i$). 
We will now prove the following crucial theorem
\begin{theorem} The following assertions are equivalent
\begin{enumerate}
\item{$\Gamma$ is a minimum norm g-inverse of $\Lambda$}.
\item{For all ${\bf y}\in\Rng (\Lambda)$, ${\bf x}=\Gamma{\bf y}$ is a solution of the equation
    $\Lambda{\bf x}={\bf y}$ with minimum norm.}
\end{enumerate}
\end{theorem} {\bf Proof.} We first prove that \ref{tbap1}
$\Rightarrow$ \ref{tbap2}. For $\Gamma$ g-inverse of $\Lambda$, one
has due to Theorem \ref{tbap2}
\begin{equation}
\begin{split}
  &\nor{\Gamma{\bf y}+(I-\Gamma\Lambda){\bf z}}^2_\pi\\=
  &[{\bf y}^\dag\Gamma^\dag+{\bf z}^\dag(I-\Lambda^\dag \Gamma^\dag)]\pi[\Gamma{\bf y}+(I-\Gamma\Lambda){\bf z}]\\=
  &\nor{\Gamma{\bf y}}_\pi^2+\nor{(I-\Gamma\Lambda){\bf z}}_\pi^2\\
  &\quad+{\bf z}^\dag(I-\Lambda^\dag \Gamma^\dag)\pi\Gamma{\bf y}+{\bf
    y}^\dag\Gamma^\dag\pi(I-\Gamma\Lambda){\bf z}.
\end{split}
\end{equation}
Since by hypothesis ${\bf y}\in\Rng(\Lambda)$, then ${\bf
  y}=\Lambda{\bf u}$ for some ${\bf u}$ in $\sK$. For a minimum norm g-inverse $\Gamma$ as in the hypothesis, due to Eq.~(\ref{minnorm}) one has
\begin{equation}
\begin{split}
  &{\bf z}^\dag(I-\Lambda^\dag \Gamma^\dag)\pi\Gamma\Lambda{\bf
    u}+{\bf u}^\dag\Lambda^\dag\Gamma^\dag\pi(I-\Gamma\Lambda){\bf z}=\\
  &{\bf z}^\dag(I-\Lambda^\dag
  \Gamma^\dag)\Lambda^\dag\Gamma^\dag\pi{\bf u}+{\bf
    u}^\dag\pi\Gamma\Lambda(I-\Gamma\Lambda){\bf z}=0.
\end{split}
\end{equation}
where the last equality is due to Eq.~(\ref{g-inv}). Finally, this
proves that
\begin{equation}
  \nor{\Gamma{\bf y}+(I-\Gamma\Lambda){\bf z}}^2_\pi=\nor{\Gamma{\bf y}}_\pi^2+\nor{(I-\Gamma\Lambda){\bf z}}_\pi^2\geq\nor{\Gamma{\bf y}}_\pi^2.
\end{equation}
namely the solution ${\bf x}=\Gamma{\bf y}$ is minimum-norm.\\
Now we prove \ref{tbap2} $\Rightarrow$ \ref{tbap1}. If ${\bf x}=\Gamma{\bf
  y}$ is a solution of $\Lambda{\bf x}=\bf y$ for all ${\bf y}\in
\Rng(\Lambda)$, by Theorem \ref{tbap1} $\Gamma$ is a g-inverse of
$\Lambda$, namely $\Lambda\Gamma\Lambda=\Lambda$. Then if $\Gamma{\bf
  y}$ is minimum norm solution of $|\Lambda{\bf x}={\bf y}|$ then due
to Theorem \ref{tbap2}
\begin{equation}
\nor{\Gamma{\bf y}}_\pi^2\leq\nor{\Gamma{\bf y}+(I-\Gamma\Lambda){\bf z}}_\pi^2
\end{equation}
for all ${\bf y}\in\Rng(\Lambda)$ and for all $\bf z$ one has
\begin{equation}
  0\leq\nor{(I-\Gamma\Lambda){\bf z}}^2_\pi+{\bf z}^\dag(I-\Lambda^\dag \Gamma^\dag)\pi\Gamma{\bf y}+{\bf
    y}^\dag\Gamma^\dag\pi(I-\Gamma\Lambda){\bf z}.
  \label{posit}
\end{equation}
Since an arbitrary ${\bf y}\in\Rng(\Lambda)$ is $\Lambda{\bf u}$ for
arbitrary $\bf u$, the second term in Eq.~\eqref{posit} becomes
\begin{equation}
\begin{split}
  &{\bf z}^\dag(I-\Lambda^\dag \Gamma^\dag)\pi\Gamma\Lambda{\bf
    u}+{\bf u}^\dag\Lambda^\dag\Gamma^\dag\pi(I-\Gamma\Lambda){\bf z}=\\
  &\quad 2\Re \! \left( {\bf z}^\dag(I-\Lambda^\dag \Gamma^\dag)\pi\Gamma\Lambda{\bf
    u} \right).
  \label{real} 
\end{split}
\end{equation}

Let us keep $\bf z$ fixed and multiply $\bf u$ by an arbitrary
$\alpha$. If the expression in Eq.~\eqref{real} is not vanishing then taking $|\alpha|$ sufficiently large, for suitable phase one can contradict the bound in Eq.~\eqref{posit}, hence $\Re \! \left( {\bf
  z}^\dag(I-\Lambda^\dag \Gamma^\dag)\pi\Gamma\Lambda{\bf u} \right) =0$ for
all $\bf u$ and $\bf z$
and by the same reasoning $\Im \! \left( {\bf z}^\dag(I-\Lambda^\dag \Gamma^\dag)\pi\Gamma\Lambda{\bf u} \right) =0$
for all ${\bf u}$ and ${\bf z}$. We can then conclude that
$(I-\Lambda^\dag
\Gamma^\dag)\pi\Gamma\Lambda=\Lambda^\dag\Gamma^\dag\pi(I-\Gamma\Lambda)=0$,
and consequently $\pi\Gamma\Lambda=\Lambda^\dag\Gamma^\dag\pi \quad \blacksquare$

Using Eq.~\eqref{minnorm}, and considering that $\Sigma_f(X)$ is the
norm of the vector of coefficients ${\bf f}[X]$ with
$\pi_{ii}=\Tr[P_i\rho_\varepsilon]$, it has been proved in
\cite{optproc} that the minimum noise is achieved by $\Gamma$
corresponding to the set of operators $D_i$ given by
\begin{equation}\label{eq:dualopt}
  D^\mathrm{opt}_i := \Delta_i-\sum_{j=1}^N\{[(I-M)\pi(I-M)]^\ddagger
  \pi M\}_{i j} \Delta_j
\end{equation}
where $\Delta_{i}$ is the set of operators corresponding to the
Moore-Penrose g-inverse $\Gamma_{mp}$ of $\Lambda$, satisfying the
properties
\begin{equation}
  \Gamma_{mp}\Lambda=\Lambda^\dag\Gamma^\dag_{mp},\ \Gamma_{mp}\Lambda\Gamma_{mp}=\Gamma_{mp},\ \Gamma^\dag_{mp}\Lambda^\dag=\Lambda\Gamma_{mp},
\end{equation}
and $M:= \Gamma_{mp}\Lambda=M^\dag=M^2$. The symbol $X^{\ddagger}$
denotes the Moore-Penrose g-inverse of $X$. It is indeed easy to
verify that
\begin{equation}
  \Gamma_{opt}:=\Gamma_{mp}-[(I-M)\pi(I-M)]^\ddagger\pi M
  \Gamma_{mp}
\end{equation}
satisfies Eq.~\eqref{minnorm}. Notice that being $\Gamma_\mathrm{opt}$
minimum norm independently of $X$, the statistical error is minimized
by the same choice $D^\mathrm{opt}_i$ for all operators $X$.

When a $N$-outcomes POVM on a $d$-dimensional Hilbert space $\sH
\thicksim {\mathbb C}^d$ is info-complete the state $\rho$ can be
written as
\begin{equation}
  \rho=\sum_{i=1}^N D_ip(i|\rho),
\end{equation}
where $D_i$ corresponds to any g-inverse $\Gamma$. It is then possible to
reconstruct any state $\rho$ using the statistics from measurements:
\begin{equation}\label{eq:reconstruction}
  \rho=\sum_{i=1}^N p(i|\rho) D_i \cong \sum_{i=1}^N \nu_i \, D^\mathrm{opt}_i,
\end{equation} 
where $\nu_i=\frac{n_i}{n_\mathrm{tot}}$ is the experimental frequency
of the $i$-th outcome, $n_i$ being the number of occurrence of the
$i$-th outcome, and $n_\mathrm{tot}=\sum_i n_i$. By the law of large
numbers we have that $\lim_{n_\mathrm{tot}\rightarrow \infty}\limits
\nu_i = p(i|\rho)$.  However, the convergence rate of $\tilde\rho$ to
$\rho$ depends on the choice of $D_i$. It turns out \cite{scott} that
the choice $D^\mathrm{opt}_i$, corresponding to $\Gamma_\mathrm{opt}$,
is the one with the fastest convergence (in average over all possible experimental outcomes) in the
Hilbert-Schmidt distance, defined as follows
\begin{equation}
  \nor{\tilde\rho-\rho_2}^2_2:=\Tr[(\tilde\rho-\rho)^2].
\end{equation}
This can be easily proved considering that the Hilbert-Schmidt
distance can be written as the sum of the variances $\delta
(|m\>\<n|)^2$, and all of the summands are minimized by the choice of
minimum-norm $\Gamma=\Gamma_\mathrm{opt}$.

\section{The Bayesian iterative procedure}\label{bay}
In this Section we describe the iterative estimation procedure based
on the update of the prior information by means of the state
reconstruction provided by experimental data. Here we provide an
algorithmic description of the procedure, that yields a
self-consistent solution:
\begin{enumerate}
\item The protocol starts with the choice of \emph{a priori} ensemble
  $\sE:=\{\rho_i,p_i\}$ (where $\rho_i$ are states and
  $p_i$ are their prior probabilities), with the corresponding density
  matrix $\rho^{(0)}:=\rho_\sE^{(0)}=\sum_ip_i\rho_i$, e.~g.  the
  one of the uniform ensemble of all pure states
  $\rho^{(0)}=I/d$.
\item Using $\rho^{(0)}$ it is possible to calculate the
  diagonal matrix with the probability of the different outcomes:
\begin{equation}
  \pi_{i j} := \delta_{i j} \Tr[P_i \rho^{(0)}]
\end{equation}   
\item Using $\pi_{i j}$ in Eq.~(\ref{eq:dualopt}) we can find the
  optimal g-inverse $\Gamma_\mathrm{opt}$ corresponding to
  $D^\mathrm{opt}_i$ associated with $\rho^{(0)}$.
\item Now the initial \emph{a priori} density matrix
  $\rho^{(0)}\equiv\rho_\sE$ will be updated as follows:
\begin{equation}
  \rho^{(1)}=\sum_{i=1}^N \nu_i \, D^\mathrm{opt}_i
\end{equation}
\item If $\rho^{(1)} \cong \rho^{(0)}$ within
  a given tolerable error $\varepsilon$ then the average input state
  is $\tilde\rho:=\rho^{(1)}$ and the
  procedure stops.
\item Otherwise after setting $\rho^{(0)}:=\rho^{(1)}$ the procedure will go back to the step 2.
\end{enumerate}

It is important to remark that at each step the matrices $\rho^{(1)}$ and $D^\mathrm{opt}_i$ are
automatically self-adjoint and normalized: $\Tr[\rho^{(1)}]=1$ since for all $i$:
$\Tr[D^\mathrm{opt}_i]=1$ \cite{optproc}, however, they are not necessarily positive.\par

This protocol in principle provides reliable state reconstructions, however, its iterative character
makes it less efficient than the one introduced in next Section, since at any iterative step one has
to calculate the Moore-Penrose g-inverse in Eq.~\eqref{eq:dualopt}, which is typically a
time-consuming operation, especially for POVM's with a large number $N$ of outcomes.

\section{The frequentist approach}\label{freq}

In this Section we introduce the second processing strategy, based on the substitution of prior
probabilities by experimental frequencies in Eq.~\eqref{minnorm}. While the previous protocol is essentially a Bayesian update, in this case the the processing relies on the law of large numbers, namely on the fact that
$\lim_{n_\mathrm{tot}\to\infty}\nu_i=p(i|\rho)$, where the limit has to be understood in
probability. We name this approach frequentist because it fits the frequentist interpretation of
probabilities as approximations of experimental frequencies, avoiding prior probabilities, which are
the signature of the Bayesian approach.

If we substitute the metric matrix $\pi$ in the Eq.
(\ref{eq:geninverse}) with the diagonal matrix of the frequencies
$\nu_{i}$, we get:
\begin{equation}\label{eq:geninverse2}
\nu \Gamma \Lambda=\Lambda^{\dagger} \Gamma^{\dagger} \nu
\end{equation}
and following the same proof as for Eq.(\ref{eq:dualopt}) we obtain
the following expression of the optimal g-inverse $\Gamma_\nu$
satisfying condition Eq.~\eqref{eq:geninverse2}, in terms of the
corresponding operators $D_i^{(\nu)}$
\begin{equation}\label{eq:dualopt2}
  D^{(\nu)}_{i} :=
  \Delta_i-\sum_{j=1}^N\{[(I-M)\nu(I-M)]^\ddagger \nu M \}_{i j}
  \Delta_j
\end{equation}
that is non linear in the outcomes frequencies due to the
Moore-Penrose g-inverse of $(I-M)\nu(I-M)$.

This protocol has the advantage that it requires only one evaluation of Moore-Penrose g-inverse, and
it is then much faster---in terms of computational resources---than the iterative one introduced in
the previous Section. However, here generally $\Tr[D^{(\nu)}_i]\neq 1$, whence in addition to
positivity of the estimated state $\tilde\rho$, also the normalization constraint is lost (but not
hermiticity). 

\section{Numerical simulations}\label{exa}
In order to test these two methods and to compare their performances with the plain un-updated
procedure some Monte Carlo simulation have ben performed. As an example, we considered the 
info-complete POVM composed by the following six elements
\begin{equation}
  P_{\pm i}=\dfrac{1}{6}(I \pm \sigma_i),
\end{equation}
$\sigma_0=I$ and $\vec\sigma=(\sigma_x,\sigma_y,\sigma_z)$ denoting the usual Pauli matrices.
The theoretical state is
\begin{equation}
  \rho =
  \begin{pmatrix}
    \frac{4}{5} & \frac{1}{7}+\frac{i}{3} \\
    \frac{1}{7}-\frac{i}{3} & \frac{1}{5}
  \end{pmatrix}=\frac12\left(I+\frac27\sigma_x-\frac23\sigma_y+\frac35\sigma_z\right).
\end{equation}

The simulation consists in 1000 experiments, each consisting in 1000
single-shot measurements, simulated by POVM events extraction
according to the theoretical probabilities $p(\pm
i|\rho):=\Tr[P_{\pm i}\rho]$. The number of iterations in the
Bayesian processing is 10.

\begin{figure}[h]
  \medskip \centerline{\includegraphics[width=7cm]{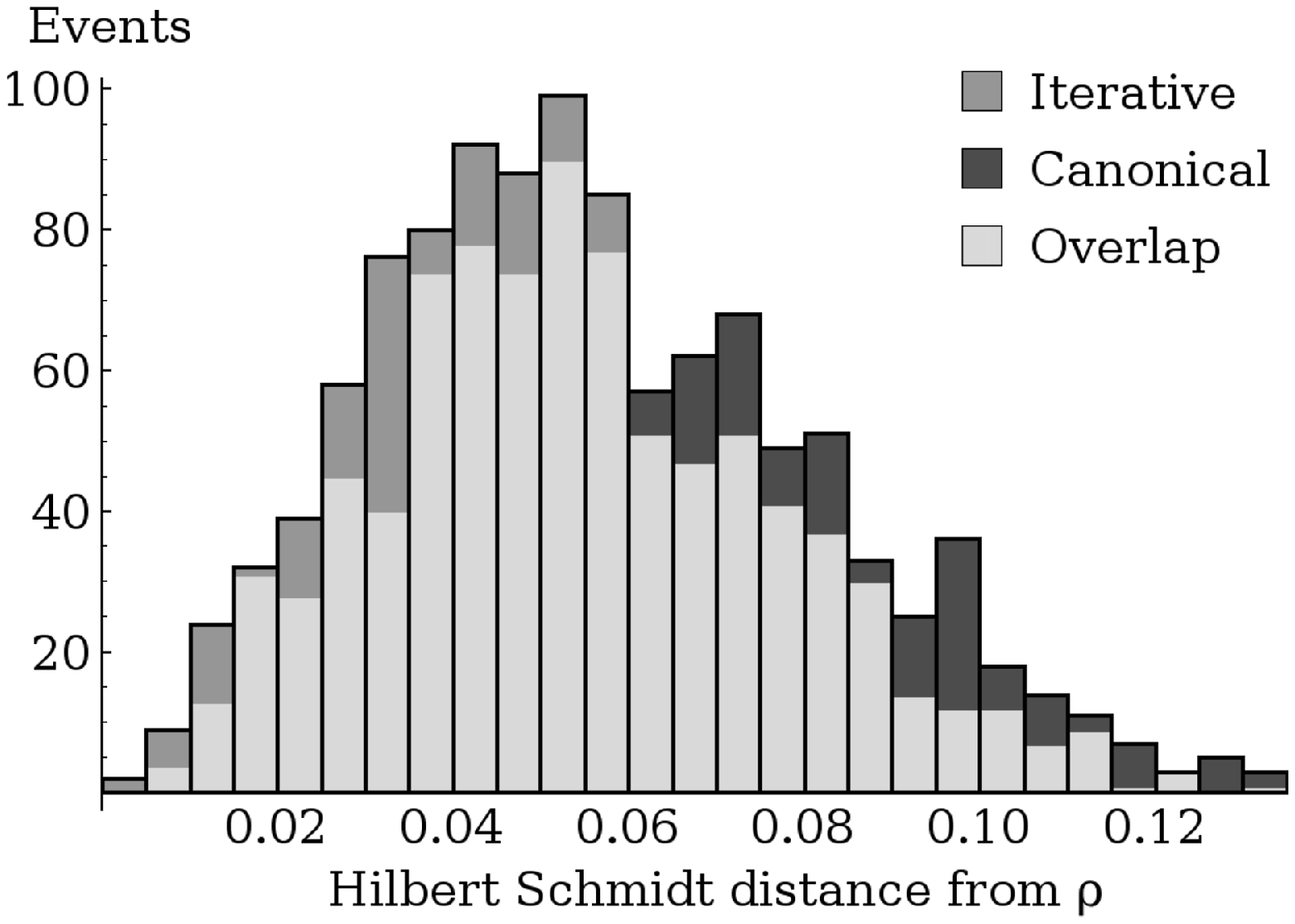} }
  \centerline{\includegraphics[width=7cm]{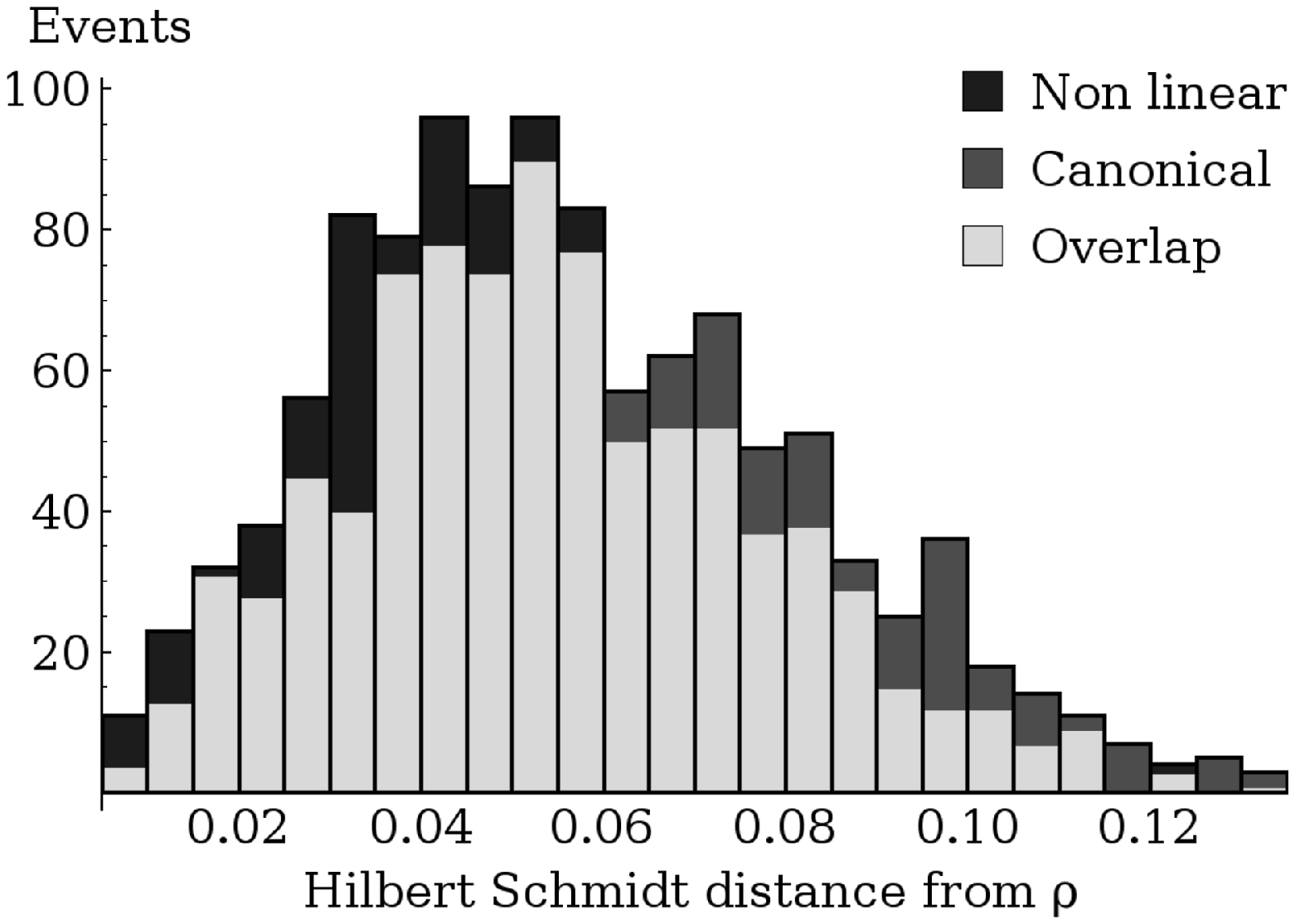} }
\caption{Histograms representing the number of experiments versus the Hilbert-Schmidt
distance of the resulting state from the theoretical one. Upper plot: the light gray bars
correspond to the Bayesian processing, the dark grey correspond to the plain processing
without updating, the white part is the overlap. Lower plot: the dark grey bars correspond to the
frequentist processing method. Both plots show a well visible shift of the histograms
corresponding to the new adaptive methods towards small errors compared to the plain processing
without update. [For other data concerning plots see text.]
\label{hists}}
\end{figure}

In Fig. \ref{hists} we show the histograms representing the number of
experiments as a function of the Hilbert-Schmidt distance of the
resulting state $\tilde\rho$ from the theoretical one $\rho$.
The plots show a well evident shift of the histograms for both new
processing methods towards small errors compared to the plain
processing without updating.  In Table \ref{tab} we summarize these
considerations by showing the average Hilbert-Schmidt distance
obtained with the three kinds of processing, along with the
corresponding variance and the relative improvement of the figure of
merit.

\begin{table}
\def\0{\hbox{\phantom{\footnotesize\rm 1}}}%.
\def\tabcolsep{1mm}
\begin{center}
  \begin{tabular}{cccccc}
    &&&&\\
    \hline
    Procedure &   $\<H.S. dist.\>$ & $\sigma$ \   &   $\Delta\left(\<H.S. dist.\>\right) $   &   $\Delta(\sigma)$ \\
    \hline
    Plain (no update) &    0.06 & 0.03  & - & - \\
    Bayesian  &    0.05 & 0.02  & -17\% & -33.3\%\\
    Frequentist  &    0.05 & 0.02  &  -17\% & -33.3\%  \\
    \hline
  \end{tabular}
\caption{Average Hilbert-Schmidt distance, variance $\sigma$ of the histogram, and relative
  improvements compared to the plain un-updated procedure of the new data-processing strategies
  presented in the paper.[For other data concerning this table see text.]
  \label{tab}}
\end{center}
\end{table}

\section{Conclusions}\label{conc}
In conclusion, we have presented two novel data-processing strategies to improve convergence of
estimation of ensemble average via info-complete measurements. The two approaches adaptively 
update the data-processing functions in a Bayesian and frequentist fashion, respectively, by
substituting the prior probabilities with experimental frequencies (frequentist) and the prior state
with the updated state (Bayesian). The two methods have been tested by numerical simulations, and
both showed improved convergence rate compared to the original plain un-updated strategy.
Clearly, further improvement is possible using both procedure together, however, this would be
an higher-order correction.\par

DFM Acknowledges financial support from CNISM, PP acknowledges
financial support from EU under STREP project CORNER.

\end{document}